\documentclass[aps,pra,reprint]{revtex4-2}

\usepackage[defaultcolor=red, authormarkup=none, final]{changes} 

\usepackage{graphicx} 
\usepackage{amsmath}
\usepackage[normalem]{ulem}
\usepackage{comment}
\usepackage{csquotes}
\usepackage{multirow}
\usepackage{xcolor}
\usepackage{setspace}
\usepackage{hyperref}
\hypersetup{colorlinks=true, allcolors=blue}
\usepackage[left=48pt,%
            right=42pt,%
            top=46pt,%
            bottom=60pt,%
            headheight=15pt,%
            headsep=10pt,%
            letterpaper,twoside]{geometry}
\usepackage[normalem]{ulem}

\usepackage{orcidlink}

\usepackage{natbib}
\usepackage{amsmath}

\usepackage{ragged2e}
\usepackage{caption}
\usepackage{subcaption}
\DeclareCaptionJustification{justified}{\justifying}
\usepackage[capitalise]{cleveref} 

\begin{document}

\title{Simulating frequency splittings and loss in Fabry-P\'erot cavities}

\author{Jonah Post\,\orcidlink{0009-0005-6663-9215}$^{1,*}$}
\author{Joep K. van den Brink$^{1}$}
\author{Martin P. van Exter\,\orcidlink{0000-0003-0839-3219}$^{1}$}

\affiliation{1. Huygens-Kamerlingh Onnes Laboratory, Leiden University, P.O. Box 9504, 2300 RA Leiden, The Netherlands}
\affiliation{* post@physics.leidenuniv.nl}

\begin{abstract}
\replaced{
Finite-element simulations of optical cavities are presented, showing frequency splittings in the resonance spectrum. These can be directly compared to and understood by recent 
theory and experiments \cite{VanExter2022FineSpectra, Koks2022ObservationStructure, Post2025OpticalModes}.
They provide strong evidence for the nonparaxial theory, but point out limitations of predicted mirror-shape corrections.
The simulations also provide model-independent predictions of modal losses for optical cavities.
}{
Finite-element simulations of optical cavities are presented, showing frequency splittings in the resonance spectrum.
These results support the theoretical framework and experimental observations presented in \cite{VanExter2022FineSpectra, Koks2022ObservationStructure, Post2025OpticalModes}. The simulated (fine) structure in the spectrum can be characterized by mirror-shape and nonparaxial effects including spin-orbit coupling.
These simulations also provide model-independent predictions of modal losses for optical cavities.}
\end{abstract}


\maketitle

\section{Introduction}
\replaced{
Optical Fabry-P\'erot microcavities are essential for cavity-QED experiments, since high finesse and small modes offer the advantage of Purcell enhancement \cite{Purcell1946SpontaneousFrequencies}.
}{
Optical Fabry-P\'erot microcavities are essential devices for cavity-QED experiments, such as the study of single emitters \cite{Haussler2021FiberCavityhBN, Herrmann2024TinVacancyMicrocavity, Bayer2023OpticalResonator}. Specifically, high finesse and small modes offer the advantage of Purcell enhancement \cite{Purcell1946SpontaneousFrequencies}.
}
Fabry-P\'erot cavities are usually described by paraxial theory predicting modes \deleted[id=textual]{that are }characterized by a longitudinal \replaced[id=textual]{mode}{quantum} number q and a transverse order N, where modes with the same $(q,N)$ labels are frequency degenerate in a rotational-symmetric cavity. 
\added{
Recent cavity experiments have used transverse modes to their advantage by coupling to them \cite{Ma2025NanophotonicElectrodynamics, Aqua2026ModeArrays}.
}
\replaced[id=textual]{However,}{But} optical micro-cavities with mirror radii $R_m$ of only a few wavelengths ($R_{\rm m}/\lambda \lesssim 100$), and cavity lengths $L<R_m$, don’t operate in the paraxial limit. Moreover, \replaced[id=textual]{micro mirrors}{micro-fabricated mirrors} come in various shapes, based on fabrication techniques \cite{Hunger2012LaserSilica, Maier2025FabricationSmoothing, Ding2026HighMicrocavities}.
This calls for a more comprehensive description taking into account nonparaxial corrections and mirror-shape effects. 

The semi-analytic theory by van Exter \textit{et al.} \cite{VanExter2022FineSpectra} takes these effects into account using perturbation theory. It predicts shifts and splittings of the eigenfrequencies relative to the paraxial model, resulting in \textit{fine structure} in the spectra of higher-order transverse modes as well as shape-induced polarization splitting of the fundamental mode. 
Although the theory has not been extensively tested yet, it has been shown to agree well with several experiments.
The fine structure has been experimentally observed by Koks \textit{et al.} \cite{Koks2022ObservationStructure}, whose cavity had close-to-spherical mirrors and operated in the strong nonparaxial regime. 
The theory also agrees well with observations of Zhang \textit{et al.} \cite{Zhang2025OpticallyArrays} in a millimeter-wave cavity, who exploited the understanding of their fine structure to avoid mode-mixing and improve finesse by two orders of magnitude.
An additional correction for an aspheric Gaussian-shaped height profile of the mirror was introduced and observed by Post \textit{et al.} \cite{Post2025OpticalModes}, whose cavity had close-to-Gaussian mirrors and operated in an intermediate regime where competition between nonparaxial, anistropic and aspheric effects was observed.

The current work aims to explore the understanding of nonparaxial and mirror-shape effects by running cavity simulations. 
\replaced{
The simulations use finite-element methods to solve the Maxwell equations with boundary conditions. They do therefore not rely on any underlying loss mechanism or model assumptions.
This numerical approach will be compared to the semi-analytical theory based on perturbative extentions of the paraxial solutions \cite{VanExter2022FineSpectra}.
Since these approaches differ vastly, it is highly unlikely that they would both be wrong by the same amount. Therefore, an agreement will indicate that both are correct.
We run these simulations for rotational-symmetric spherical and Gaussian-shaped mirrors.
}{
The simulations are based on finite-element methods, while the theory is based on perturbation theory. If simulations and theory agree, then they are probably both correct, because it would be highly unlikely that they would both be wrong by the same amount.
We run these simulations for rotational symmetric perfectly spherical and Gaussian shaped mirrors.
}
\added{The setup is available online \cite{Post2026Fabry-PerotProfiles}.}

In practice, experimental scientists want to couple efficiently to cavity-modes and limit their cavity loss. An understanding of the dominant mode-shaping effects in their cavity is then essential to optimize their alignment and cavity geometry in order to achieve this.
In \cite{Post2025OpticalModes} we already provided a recipe to characterize the dominant effects in a cavity. In addition, researchers can now use the presented simulation setup to supplement their measurements. With the combined results they can have a robust understanding of all beyond-paraxial effects in their cavity, allowing them to, for instance, exploit the polarization splitting of the fundamental mode and avoid any unwanted degeneracies, which could help to circumvent avoidable losses.
\deleted{Simulations presented in this work are finite-element simulations and therefore do not rely on any underlying loss mechanism or model assumptions.
The setup is available online \cite{Post2026Fabry-PerotProfiles}.}

\section{Theory}
Resonances in an optical cavity are determined by the criteria that the field should reproduce itself after each roundtrip and that the roundtrip phase should be a multiple of $2\pi$, \replaced[id=textual]{i.e.}{e.g.} they should be eigenmodes of the roundtrip operator. For a \added[id=textual]{plano-concave} cavity of length $L$ \deleted[id=textual]{with ideal mirrors}, the (dimensionless) eigenfrequencies are given by
\begin{equation} \label{eq:resonance_conditions}
    \widetilde{f}_j \equiv \frac{f_j}{f_{\rm FSR}} = \frac{2L}{\lambda_j} = q + (N+1)\frac{\chi_0}{\pi}+\Delta \widetilde{f}_j\,,
\end{equation}
\replaced[id=textual]{where $q$ is the longitudinal mode number, $N$ is the transverse mode number and $f_{\rm FSR} = c/2L$ is the free spectral range (FSR). }
{where $q$ is the longitudinal mode number and $N$ is the transverse mode number. The absolute frequency difference between two consecutive longitudinal modes $f_{\rm FSR} = c/2L$ is known as the free spectral range (FSR).}
The first term describes the resonances of a planar-planar cavity. The plano-concave shape introduces transversal confinement, leading to the second term via the paraxial Gouy phase $\chi_0=\arcsin{\sqrt{L/R_{\rm m}}}$ for a spherical mirror with radius of curvature $R_{\rm m}$.
This paper focuses on the final term $\Delta \widetilde{f}_j = \Delta \widetilde{f}_{j, \text{non}} + \Delta \widetilde{f}_{j, \text{asp}}$, which describes shifts in frequencies with respect to the paraxial theory\deleted[id=textual]{,}\added[id=textual]{. For rotational-symmetric cavities, theory predicts that these shifts are} caused by
(i) a nonparaxial correction\deleted[id=textual]{;} and (ii) an aspheric mirror-shape correction.

The nonparaxial correction is given by \cite{VanExter2022FineSpectra}
\begin{equation} \label{eq:nonparcorrection}
\begin{split}
    \Delta \widetilde{f}_{j,\text{non}} & = \frac{1}{2\pi k R_{\rm m}} \left[-\ell \cdot s - \frac{3}{8}\ell^2 + \frac{3}{8} \langle \ell^2\rangle_{\ell} - \frac{1}{2}\right]\,,
\end{split}
\end{equation}
where the azimuthal mode number $\ell \in \{ -N, -N+2, ..., N\}$ quantizes orbital angular momentum, commonly used for the Laguerre-Gauss (LG) modes, and the photon spin $s \in \{ 1, -1\}$ quantizes circular polarization. 
The third term includes an average over all \deleted[id=textual]{these} values of $\ell$\replaced[id=textual]{ via}{, i.e. one could also write} $\langle \ell^2\rangle_{\ell} = N (N+2)/3$.
The spin-orbit coupling $\ell \cdot s$ is associated with the vectorial properties of the electric field \cite{VanExter2022FineSpectra}\deleted[id=textual]{. 
The absolute sign of $\ell$ is irrelevant since}\added[id=textual]{, where} only the relative sign $\ell \cdot s$ matters. Each $(q,N)$ group is thus split into $N+1$ resonances.
We will denote modes by their azimuthal mode number $\left|\ell\right| >0$, appended by $A$ for $\ell \cdot s <0$ or $B$ for $\ell \cdot s >0$; a naming convention adopted from Yu and Luk \cite{Yu1983High-orderResonator}. \replaced[id=textual]{\eqref{eq:nonparcorrection} predicts}{We see from \eqref{eq:nonparcorrection}} that $\ell B$ modes have a lower frequency than $\ell A$ modes. 
The invariance of \eqref{eq:nonparcorrection} under $(\ell, s )\to (-\ell, -s)$ leaves a two-fold degeneracy, which is distinguishable in the polarization profiles $\vec{E}(x,y)$. 
The profiles of these two modes have opposite parity upon $x$-reflection $\vec{E}(x,y) \to \vec{E}(-x,y)$ and will be labeled $+$ (even) and $-$ (odd).

The aspheric correction originates from the deviation from a spherical mirror shape as defined by the height profile:
\begin{equation}
    z_{\text{sphere}} (r,\theta) = L+ \sqrt{R_{\rm m}^2-r^2}-R_{\rm m} \,.
\end{equation}
A common shape, as a result of CO$_2$ laser ablation as a technique for micro-mirror fabrication \replaced{\cite{Hunger2012LaserSilica}}{\cite{Najer2017FabricationCurvature}}, is the Gaussian mirror:
\begin{equation}\label{eq:mirrorshape_gauss}
    z_{\text{Gauss}} (r,\theta)= L-h+h\exp{\left( -\frac{r^2}{2hR_{\rm m}} \right) } \,,
\end{equation}
where $h$ is the mirror depth.
The aspheric correction for a Gaussian shaped mirror is \cite{Post2025OpticalModes}
\begin{equation}\label{eq:aspcorrection}
    \Delta \widetilde{f}_{j, \text{asp}} = \frac{G}{2\pi k R_{\rm m}} \left[\frac{3}{8}\ell^2 - \frac{3}{8} \langle \ell^2\rangle_{\ell} - g_{\text{asp}}(N)\right]\,,
\end{equation}
where $g_{\text{asp}}(N) = (N^2+2N+\frac{3}{2})$ and $G = \frac{L}{3h} \left( \frac{R_{\rm m}}{R_{\rm m}-L} \right)$ \cite{Koks2022ObservationStructure, Post2025OpticalModes}.
\added{
\eqref{eq:aspcorrection} is an approximate result that is based on a Taylor expansion of the mirror shape, \eqref{eq:mirrorshape_gauss}, in powers of $\left( r^2 / h R_{\rm m}\right)$, only up to order $(r^2 / h R_{\rm m} )^2$. This expansion works well for low-$N$ modes but is 'disappointingly bad` \cite{Post2025OpticalModes} for higher-order modes in larger cavities, where $\left< r^2 \right>/ h R_{\rm m}$ becomes too large.
}


The cavity is thus characterized by 3-4 length scales: (i) optical wavelength $\lambda$, (ii) cavity length $L$, (iii) mirror radius of curvature $R_{\rm m}$, and in the case of Gaussian geometry (iv) mirror depth $h$. Because Maxwell's equations are scale invariant, the absolute lengths do not matter, only their ratios. All results will therefore be presented in dimensionless ratio's. 

An important characteristic of an optical resonator, if to be used for QED-experiments or applications, is the loss, or alternatively the finesse. 
Since the simulation uses ideal mirrors that reflect $100\%$, our simulated loss is \replaced{only caused by power leaking out in the transverse direction}{`mode mismatching loss'}.
The imaginary part of the simulated complex eigenfrequencies characterizes the exponential decay of the electric field. 
\replaced[id=textual]{We express the simulated loss}{This simulated loss can thus be expressed} in finesse $F$ as:
\begin{equation} \label{eq:loss}
    \frac{1}{F} =  \frac{A}{2\pi} = 2\frac{\operatorname{Im}(f_j)}{f_{\rm FSR}}\,,
\end{equation}
where $A$ is the relative intensity loss per roundtrip.


\section{Simulations}
To simulate microcavities, the Maxwell equations with boundary conditions have to be solved to find resonance frequencies.
The simulations are performed in COMSOL Multiphysics \cite{COMSOLMultiphysics2018COMSOLMultiphysics} using the Eigenfrequency Solver of the Wave Optics Module. 

We first define the geometry with a concave top and planar bottom mirror surrounded by a cylindrical box that includes a cylindrical perfectly-matching layer \added{(PML)}. \added{The PML absorbs the field leaking out of the cavity, but is placed far enough not to influence the geometric loss rate of this field.} 
\replaced{The geometry is then discretized using a tetrahedral mesh, using element sizes from $\lambda/60$ in the longitudinal direction to $\lambda/3$ is the outer regions of the transverse direction.}{This geometry is then discretized using a tetrahedral mesh. We typically choose the one-but-finest mesh, called 'Extra fine', which we then make a little bit finer.}
These meshes yield equations with 20,000-120,00 domain elements and 200,000-1,200,000 degrees of freedom. 
The computation times needed to obtain all relevant resonance frequencies was 1-10 minutes on an ordinary PC (Intel Core i5-14500 CPU, 4 × 32 GB RAM).
We have also used the finest preset mesh size, called 'Extremely fine' \added{(element sizes $\lambda/400$ - $\lambda/4$)}, to verify the quality of some results and
conclude that our mesh was fine enough to determine the frequencies shifts $\Delta \tilde{f}$ to $ 10^{-4}$. 
The simulations provide resonance frequencies alongside with full three-dimensional (polarization-resolved) mode profiles. By visual inspection of cross sections, associated mode numbers can reliably be determined.

Two geometries were explored. The first cavity has a spherical mirror, for which the only beyond-paraxial correction is the nonparaxial term, \eqref{eq:nonparcorrection}. The second geometry has a Gaussian mirror, for which also the aspheric correction, \eqref{eq:aspcorrection}, becomes relevant.
These cavities were primarily analyzed for $R_{\rm m} = 150$ cm for wavelengths $\lambda \sim 11-12$ cm, such that they were similar to those in \cite{Koks2022ObservationStructure, Zhang2025OpticallyArrays, Post2025OpticalModes}, where typically $10<R_{\rm m}/\lambda <30$. 
Furthermore $h=6$ cm was chosen for the Gaussian cavity such that $2h/\lambda \sim 1$, close to the experimental conditions in \cite{Post2025OpticalModes}.

\section{Results}
\subsection{Results Spherical Mirror}
\begin{figure}[t]
     \centering
     \begin{subfigure}{\linewidth}
        \centering
        \includegraphics[width=\linewidth]{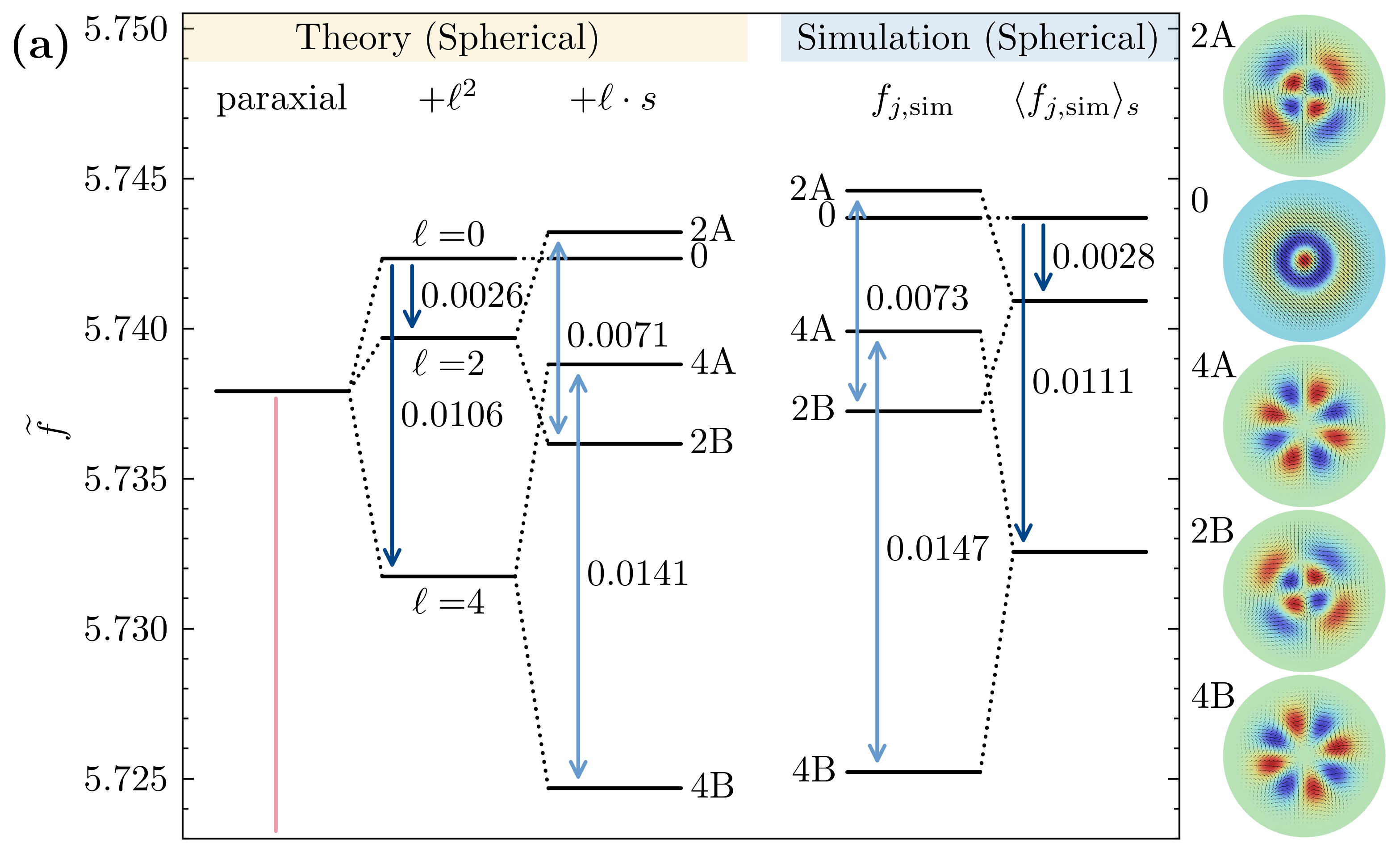}
        \phantomsubcaption
        \label{fig:resonances_sphere}
     \end{subfigure}
     \hfill
     \begin{subfigure}{\linewidth}
        \centering
        \includegraphics[width=\linewidth]{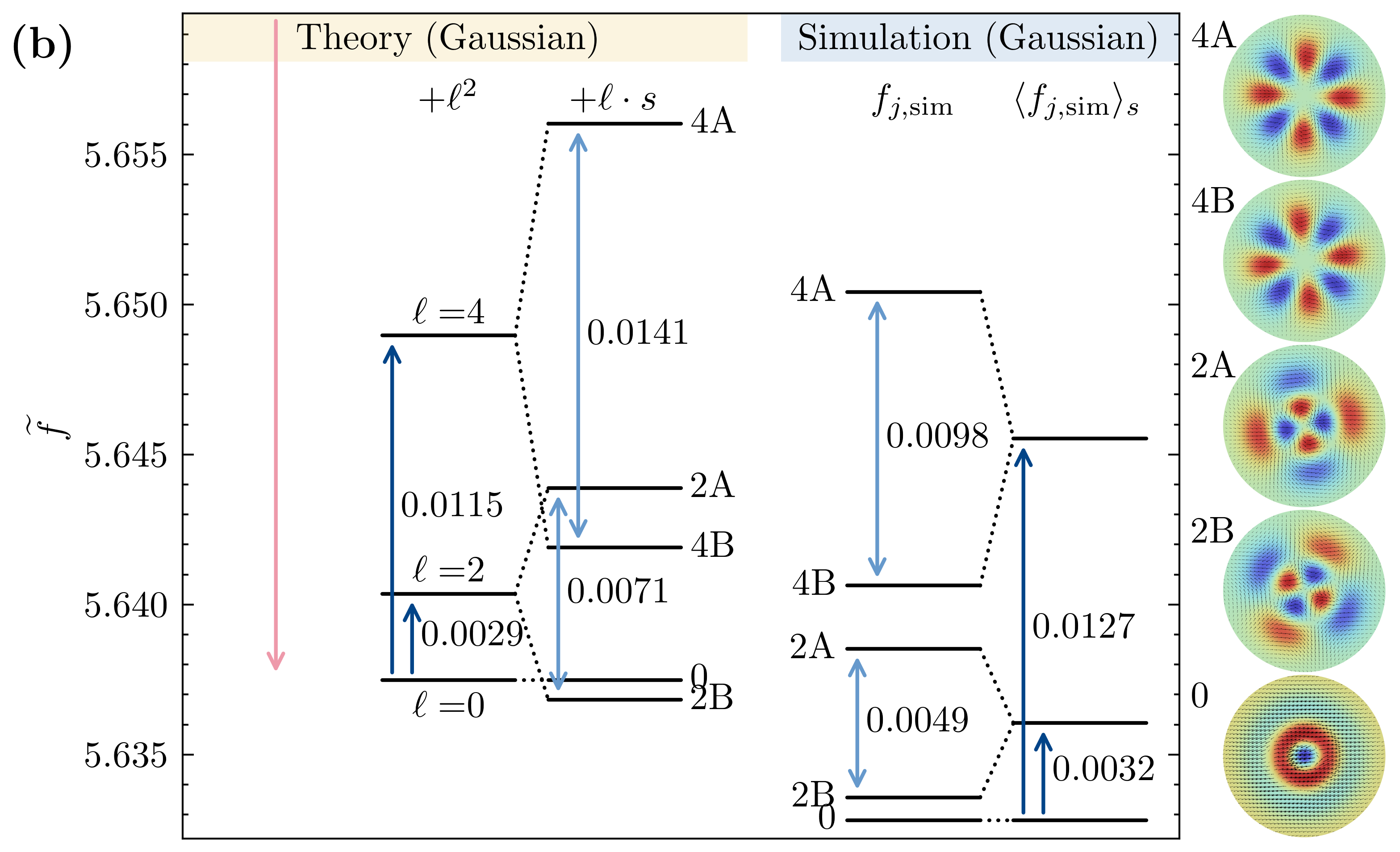}
        \phantomsubcaption
        \label{fig:resonances_gauss}
     \end{subfigure}
        \caption{Level scheme with fine structure of the $q=5, N=4$ resonance group for (a) spherical and (b) Gaussian cavities at $L/R_{\rm m} = 0.2$.
        The theoretical predictions are shown on the left. The computed eigenfrequencies $f_{j,\text{sim}}$ are shown on the right. The $\ell^2$ splitting (dark blue arrows) and spin-orbit coupling (light blue arrows) can be compared. The images on the far right show the simulated modal field profiles.}
        \label{fig:three graphs}
\end{figure}

The predicted and computed eigenvalues of the spherical cavity are shown in Fig. \ref{fig:resonances_sphere}.
Theoretical predictions on the left show paraxial results, shifts due to the $\ell^2$ term (dark blue arrows), and the symmetric splittings due to the $\ell \cdot s$ term (light blue arrows).
The simulated frequencies $f_{j,\text{sim}}$ on the right exhibit a similar division.
The simulation accurately produces the predicted $N+1$ eigenmodes within each $(q,N)$ group. The frequency difference between $A$-$B$ pairs (light blue arrows) characterizing the spin-orbit coupling agrees with predictions $<4\%$. The pairwise average $\langle f_{j,\text{sim}}\rangle _{s}$ is calculated, of which the frequency shifts (dark blue arrows) can be compared to the predicted $\ell^2$ splitting, these agree with the prediction $<8\%$. 
For both effects the simulated splitting is slightly larger than the predicted splitting. The same conclusion is drawn for all analyzed $N$-groups, which were $N=0$ up to $N=6$.
The images on the far right show the simulated cross sectional mode profiles, $\operatorname{Re}[E(x,y)]$.

The \replaced[id=textual]{simulations confirm}{results also include} the predicted two-fold degeneracy visible in the polarization profiles (in Fig. \ref{fig:resonances_sphere} only one profile of each pair is shown). This degeneracy serves as an internal quality check of the simulation. 
Quite intriguingly, we observe a small but systematic frequency splitting between the $1A+$ and $1A-$ modes. Although this type of \textit{hyperfine} structure might be expected in experiments with Bragg mirrors \cite{VanExter2022FineSpectra}, this remains unexplained for the perfect mirrors used in COMSOL.

\replaced{
The whole $(q,N)$ group of simulated frequencies is offset by $1\times 10^{-3}$ in $\widetilde{f}$ relative to the predicted group of frequencies.
}{There is also a small offset discrepancy of about $1\times 10^{-3}$ in $\widetilde{f}$.}
To emphasize accuracy, this \added{offset discrepancy} should be compared to the absolute offset of the $(q,N)$ group, given by the paraxial Gouy phase
in this case $5\chi_0/\pi \approx 0.7$. Even the offset is thus accurate to  $\sim 0.1\%$.

\subsection{Results Gaussian Mirror}

In Fig. \ref{fig:resonances_gauss} a comparison between theory and simulated Gaussian cavity is depicted.
Compared to the spherical cavities, all resonances are found at lower frequencies (red arrow) and the $\ell^2$ shifts have a positive instead of negative sign. Both changes are also predicted by theory. 
\replaced[id=textual]{
The first observation originates from the reduced transverse confinement in Gaussian mirrors, described by the two last terms in \eqref{eq:aspcorrection}. The second observation is expected for $G>1$, which holds here.
}{
Reduced transverse confinement in Gaussian mirrors results in an overall shift, accounted for by the two last terms in \eqref{eq:aspcorrection}, explaining the first observation. The latter observation is expected for $G>1$, which holds here.}

For Gaussian mirrors, agreement between theory and simulation is reasonable, but far from perfect for predicting exact frequencies. 
We observe three quantitative differences: (i) the $\ell=0$ mode is found at a slightly lower frequency, (ii) simulated $\ell^2$ \replaced[id=textual]{splittings are}{splitting is} larger by $10\%$, and (iii) simulated $\ell \cdot s$ \replaced[id=textual]{splittings are}{splitting is} smaller by $30\%$ than predicted. 
Discrepancy (i), being $5\times 10^{-3}$ in $\widetilde{f}$, again must be compared to the paraxial shift set by the Gouy phase $5\chi_0/\pi \approx 0.7$ or to the additional group shift caused by $\langle \widetilde{f}_{j,\text{asp}}\rangle _{\ell} \approx -0.1$. The offset is then in good agreement, accurate to a few percent.
\added{
When studying results of all $N$ groups, we note the predicted $\ell\cdot s$ splittings fails and that the simulated splittings are smaller, especially for larger $N$, resulting in difference (iii) being more excessive for larger $N$.}
\deleted{Difference (ii) might be caused by the theory only accounting for the Gaussian mirror profile in an expansion up to order $r^4$. Extended mode profiles might `feel' more of the wings of the mirror, and therefore higher-order Taylor terms would be required to achieve a higher accuracy.
Discrepancy (iii) results in a different mode order for this particular simulation. 
When studying results of all $N$ groups, we note that the expected proportionality relation $\ell \cdot s \propto \ell$ fails, especially for larger $N$, resulting in difference (iii) being more excessive for larger $N$. Theory predicts equal $l\cdot s$ coupling for both spherical and Gaussian cavities while computed values are smaller for Gaussian cavities. We attribute this discrepancy to approximations in the theory.}

\begin{figure}
\centering
\includegraphics[width=\linewidth]{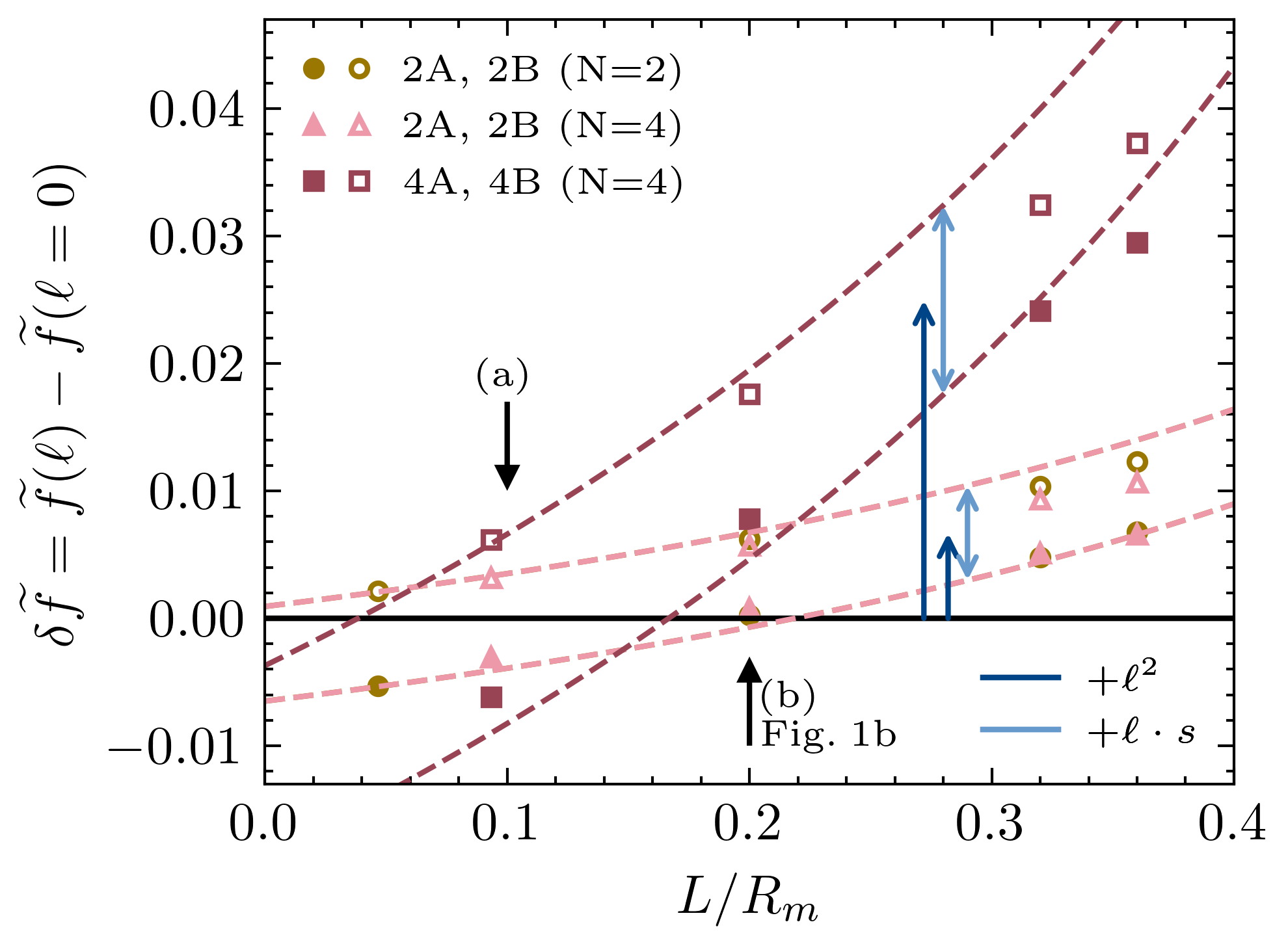}
\caption{
The $\ell$-dependent splitting $\delta \widetilde{f} = \widetilde{f}(\ell) - \widetilde{f}(\ell=0)$ as a function of cavity length for a Gaussian mirror.
The dashed curves present the theoretical predictions for $kR_{\rm m}\approx2\pi R_{\rm m}/\lambda_{\text{par}}$ 
with $R_{\rm m} / \lambda_{\text{par}} = 13.6$ and $2h/\lambda_{\rm p ar} = 1.09$, both set by the simulation geometry, without any fitting parameter.
}
\label{fig:N4}
\end{figure}

Resonance spectra were also simulated for other $L/R_{\rm m}$ ratios. Fig. \ref{fig:N4} shows these results for the six $\ell \neq 0$ modes in the $N=2$ and $N=4$ groups, relative to the $\ell=0$ mode.
Data are computed for the same $R_{\rm m}$ at different cavity lengths where we have chosen $q$ such that  $\lambda_{\rm par}/R_{\rm m}$, with $\lambda_{\rm par}= L/(q+\frac{\chi_0}{\pi}) \approx11\ \text{cm}$, remained (almost) constant while $L/R_{\rm m}$ varied. Dashed curves present theoretical prediction for $R_{\rm m} / \lambda_{\text{par}} = 13.6$. 

The $\ell^2$ behavior \replaced[id=textual]{is}{can be} observed as the deviation of the $A$-$B$ pairs from the $\ell=0$ resonance \deleted[id=textual]{as }indicated by the dark blue arrow.
The averages of the computed $A$-$B$ pairs lie closer to the $\ell=0$ resonance than that of the corresponding theoretical curves, therefore the aforementioned discrepancy (ii) also holds for larger cavity lengths. 
The $\ell \cdot s$ behavior can be observed in the splitting of each $A$-$B$ pair visualized by the light blue arrow.
The simulated pairwise splittings are smaller than theoretical predictions, thereby extending discrepancy (iii) for all cavity lengths.

At small cavity lengths, most eigenfrequencies lie at lower frequencies than the $\ell=0$ resonance. This shows that nonparaxial effects dominate since the $\ell^2$ term in the nonparaxial correction in \eqref{eq:nonparcorrection} has a negative sign.
As the cavity length increases, the $\ell^2>0$ resonances shift to higher frequencies, eventually above the $\ell=0$ resonance. Here the aspheric effects take over since $G$ in \eqref{eq:aspcorrection} increases with cavity length.
We thus observe the predicted length-dependent competition between the nonparaxial and Gaussian corrections.
An intriguing case then appears around $L/R_{\rm m} = 0.1$, indicated by arrow (a), where we find the $\ell=2$ and $\ell=4$ resonances spread symmetrically around the $\ell=0$ resonance. 
This is the case where $G\approx1$ and the $\ell^2$ term disappears, \replaced[id=textual]{i.e.}{e.g.} the aspheric correction cancels out the average non-paraxial shifts.
The $L/R_{\rm m} = 0.2$ case of Fig. \ref{fig:resonances_gauss}, indicated by arrow (b) in Fig. \ref{fig:N4}, is interesting since several frequencies lie close together. This makes the mode order very susceptible to small discrepancies between theory and simulation, explaining \replaced{how discrepancy (iii) leads to an}{the aforementioned} observed difference in mode order \added{for the particular case of Fig. \ref{fig:resonances_gauss}}.

\begin{figure}
\centering
\includegraphics[width=\linewidth]{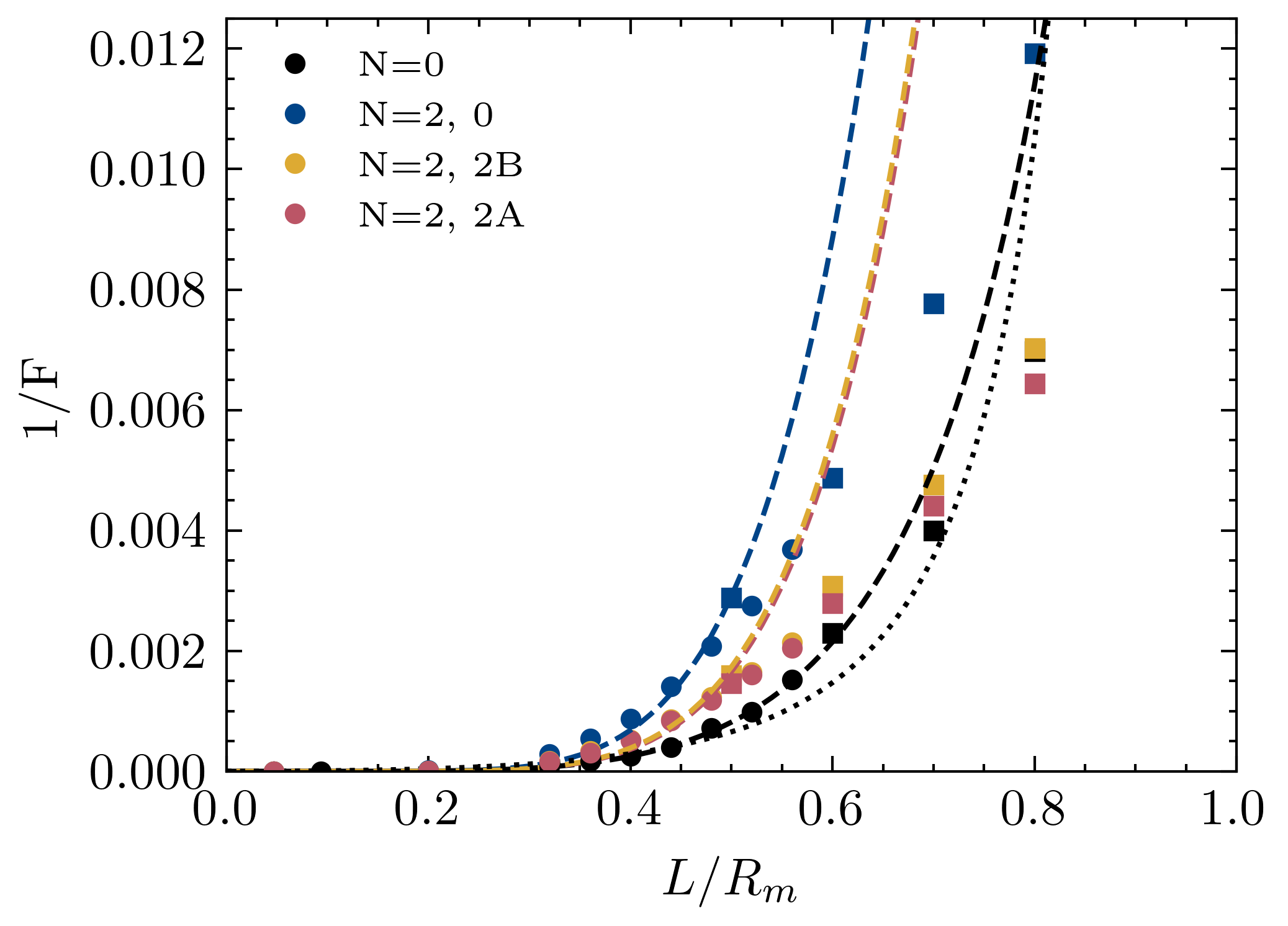}
\caption{Loss, plotted as 1/finesse as a function of cavity length for the N=0 and N=2 resonances for Gaussian cavities. The circular markers present data for $R_{\rm m}=150 \rm\ cm$ while the square markers present additional data for $R_{\rm m}=60\rm\ cm$ (these were computed since smaller $R_{\rm m}$ requires less computation time).
The dashed and dotted curves present the deflection loss and scattering loss models proposed in \cite{Post2025OpticalModes} respectively.
}
\label{fig:Losses}
\end{figure}

The imaginary part of the simulated eigenfrequencies yields the modal losses according to \eqref{eq:loss}.
Fig. \ref{fig:Losses} shows the loss as inverse finesse as a function of cavity length.  
We observe that the losses increase with cavity length. 
Modes with higher $N$ and modes with lower $\ell$ (for equal $N$) experience higher losses.
These modes have more power in the outer areas of their mode profiles and are thus more prone to \replaced{power leakage out of the cavity}{scattering and clipping losses}, explaining the above observations.
We also observe that the 2B modes are slightly more lossy than 2A modes, which we cannot yet explain.
Both the deflection loss and scattering loss models proposed in \cite{Post2025OpticalModes} have been fit on a logarithmic scale to the $L/R_{\rm m}>0.2$ data, with $hk$ as fit parameter.
The deflection loss seems to fit the loss for $N=0$ well, but is inaccurate for the $N=2$ modes. The scattering loss does not model the data. 

Cavity loss can also be simulated with resonant state expansion models, such as described by Kleckner \textit{et al.} \cite{Kleckner2010Diffraction-limitedCavities}. 
Such cavity computations were performed, among others, by Benedikter \textit{et al.} 
to simulate losses due to mode-mixing in Gaussian cavities \cite{Benedikter2015Transverse-modeMicrocavities} and due to surface micro roughness and local curvature gradients of the mirrors \cite{Benedikter2019Transverse-modeMicroscopy}. 
The simulations presented in the current work are finite-element simulations for smooth mirrors. They do not rely on any underlying loss mechanism or model assumptions.
Additionally, these simulations provide $\vec{E}(\vec{r})$ throughout the cavity. By visual inspection, one can thus take a `look under the hood' to identify loss mechanisms, something that is not be possible in experiments. 
This inspection shows \replaced[id=textual]{for instance}{, among other things,} that some light leaks out of the cavity in the transverse direction, as \added[id=textual]{experimentally} observed in \cite{Post2025OpticalModes}.

\section{Discussion}
\deleted[id=textual]{To conclude, the}\added[id=textual]{The} finite-element simulations produce \added[id=textual]{spectral} fine structure with the characteristic $\ell^2$ and $\ell \cdot s$ behavior predicted by the combined nonparaxial and Gaussian corrections\deleted[id=textual]{\cite{VanExter2022FineSpectra, Post2025OpticalModes}}.
For spherical mirrors, an accurate match is observed.
\deleted[id=textual]{For Gaussian mirrors, the modified $\ell^2$ and $\ell \cdot s$ splittings and mode competition are observed. However, there remains a slight quantitative discrepancy in predicted mode splittings, in particular for the spin-orbit coupling term, where $\left(\ell\cdot s / 2\pi k R_{\rm m} \right)_{\rm sim} < \left(\ell\cdot s / 2\pi k R_{\rm m} \right)_{\rm theory}$}
\added{
For Gaussian mirrors, modified $\ell^2$ and $\ell \cdot s$ splittings and mode competition are observed. However, discrepancies (ii) and (iii) show that the simulated splittings are smaller than predicted by the theory, specifically for higher $N$ modes in larger cavities.
Our simulations thus suggest that the current perturbation theory for Gaussian cavities is incomplete and requires improvements. We discuss two limitations which could be improved.
First, as discussed in \cite{Post2025OpticalModes}, the theory only accounts for the Gaussian mirror profile up to order  $(r^2 / h R_{\rm m} )^2$, see \eqref{eq:aspcorrection}.
For our case of Fig. \ref{fig:resonances_gauss}, where N=4, $L/R_{\rm m}=0.2$ and $2h/\lambda\approx 1$, we find $\left< r^2 \right>/ h R_{\rm m} \approx (N+1)\sqrt{L(R_{\rm m} - L)}/k h R_{\rm m} \approx 0.6$, which is only slightly smaller than 1, so higher-order Taylor terms are required to achieve higher accuracy.
A second limitation of the theory lies in the distinction between the mirror radius and the wavefront radius. In \eqref{eq:nonparcorrection} and \eqref{eq:aspcorrection} they are quantified by a single parameter $R_{\rm m}$ and are therefore assumed to be equal and fixed, which only holds for spherical mirrors. The Gaussian mirror shape, however, provides less transverse confinement. One could introduce an effective radius of curvature $R_{\rm eff} \geq R_{\rm m}$, as an intensity weighted average, increasing with beam size on the mirror, to explain the smaller magnitude of the $l\cdot s$ coupling as observed, see Appendix B in ref. \cite{Post2025OpticalModes} for an extensive discussion.
}

While the paraxial theory correctly predicts shifts only up to $\delta f / f \sim 10^{-1}$, the simulations and full theory \added{including the non-paraxial and aspheric corrections} are in agreement up to $\delta f / f \sim 10^{-4}$\added[id=textual]{, which is three orders of magnitude more accurate.} \deleted[id=textual]{We conclude that the full theory can accurately predict frequency shifts and splittings up to three orders of magnitude more accurate than the paraxial theory.}

The simulations also provide data on cavity losses, which are reliable since no underlying loss-mechanism is assumed.
The visualization of $\vec{E}(\vec{r})$ throughout the 3D geometry potentially allows one to pinpoint underlying loss mechanisms.

We emphasize that these simulations are computationally inexpensive. It is feasible for experimentalists, designing or working with microcavities, to run these simulations to understand which mode-shaping effects are relevant in their operational regime. They can exploit this to avoid mode degeneracies and find high-finesse operating regimes. The simulations can be extended to include the mirror anisotropy as a third correction.

\section{Acknowledgments}
This publication is part of the project “Optical microcavities and 2D quantum emitters” with File No. NGF.1623.23.015 of the research programme NGF - Quantum Delta NL Quantum Technologie 2023, which is (partly) financed by the Dutch Research Council (NWO).

\bibliography{references}

\end{document}